# The complex kinetics of the ice VI to ice XV hydrogen ordering phase transition

Jacob J. Shephard[a] and Christoph G. Salzmann[*,a]

[a] *Department of Chemistry, University College London, 20 Gordon Street, London WC1H 0AJ, United Kingdom*

**\*** e-mail: c.salzmann@ucl.ac.uk, telephone: +44 (0)20 7679 8864

**Abstract**

The reversible phase transition from hydrochloric-acid-doped ice VI to its hydrogen-ordered counterpart ice XV is followed using differential scanning calorimetry. Upon cooling at ambient pressure fast hydrogen ordering is observed at first followed by a slower process which manifests as a tail to the initial sharp exotherm. The residual hydrogen disorder in $H_2O$ and $D_2O$ ice XV is determined as a function of the cooling rate. We conclude that it will be difficult to obtain fully hydrogen-ordered ice XV by cooling at ambient pressure. Our new experimental findings are discussed in the context of recent computational work on ice XV.

**Keywords**

ice, hydrogen ordering, residual entropy, (anti)ferroelectrics, hydrogen bonding

## 1. Introduction

Ice VI can be crystallised from liquid water in the 0.6 to 2.2 GPa pressure range.[1] This tetragonal phase of ice with $P4_2/nmc$ space group symmetry consists of two independent and interpenetrating hydrogen-bonded networks.[2,3] The basic building blocks of the networks are 'cage-like' $(H_2O)_6$ clusters[4,5] which share corners in the crystallographic $c$ direction and are hydrogen-bonded to one another in the $a$ and $b$ directions. Like all other known phases of ice that can be crystallised from the liquid, ice VI is hydrogen disordered, *i.e.* it displays orientational disorder of the hydrogen-bonded water molecules.[2,3] The hydrogen disorder is in principle expected to disappear at low temperatures in line with the 3$^{rd}$ law of thermodynamics. Yet, upon cooling pure ice VI the orientational disorder is frozen-in resulting in an orientational glass.[3,6] This is due to the highly cooperative nature of hydrogen ordering processes in ice which rely on the presence and mobility of point defects that locally lift the constraints imposed by hydrogen bonding on the orientations of the water molecules. We have shown that doping with hydrochloric acid (HCl) enhances molecular reorientation in ice VI which then facilitates hydrogen ordering at low temperatures and consequently the phase transition to hydrogen-ordered ice XV.[1,7,8] While there is only one fully hydrogen-disordered state a multitude of candidate hydrogen-ordered structures can be envisaged. Depending on the relative orientations of the water molecules these can be either antiferroelectric or ferroelectric. With the same unit cell size as ice VI there are three different ways in which the hexameric units can become hydrogen-ordered. These are labelled *A*, *B* and *C*-type networks and are shown in Figure 1(a-c). The *C*-type network is considerably more polar than the *A* and *B* networks.[1,7] The symmetry relationship between the two hydrogen-ordered networks in ice XV then determines the space group symmetry. Cancellation of the polarities of the networks are found in *P*-1 whereas ferroelectric structures have either *P*1, *P*2$_1$, *Pn* or *Cc* space group symmetry.[7] Our neutron diffraction analysis of D$_2$O ice XV has shown a clear preference for *C*-type networks. The best fit to the neutron diffraction pattern was obtained with the *P*-1 structure. A ferroelectric $P2_1$ structure was found to give a good fit as well. Yet, it was discarded on the basis that Landau theory does not permit a one-step phase transition to $P2_1$ as it is possible for *P*-1.[7]

The various hydrogen-ordered candidate structures of ice XV and other phases of ice have been used extensively as benchmark structures for the computer models of water.[1,9-17] The expectation is that atomistic quantum mechanical or classical models should identify an experimental hydrogen-ordered structure as the lowest energy structure. Ice XV has been a particular challenge in this respect. Early computational work suggested, in contrast with the experimental findings, that the lowest energy structure is ferroelectric with *A*-type networks and *Cc* space group symmetry.[11,12] However, a more recent study using fragment-based second-order perturbation and coupled cluster theory identified the *P*-1 structure with *C*-type networks as the lowest energy structure.[16] This result was later questioned by Del Ben *et al.* on the basis of fully periodic 2$^{nd}$ order Møller-Plesset perturbation theory and random phase approximation approaches which reconfirmed the earlier *Cc* structure as the energetic ground state.[17] Experimentally, we have shown that the *Cc* structure is not consistent with the experimental Raman spectrum of ice XV.[18] Furthermore, a *Cc* structure can be excluded due to the presence of a Bragg peak at 1.93 Å which is absent in *Cc*.[7] The disagreement between experiment and theory is therefore still not resolved.

Here we follow the hydrogen ordering phase transitions of HCl/DCl-doped ice VI samples at ambient pressure using differential scanning calorimetry (DSC). The aim is to gain new insights into the kinetics of the hydrogen ordering phase transition and to quantify the residual hydrogen disorder in $H_2O$ and $D_2O$ ice XV.

## 2. Experimental Methods

For the preparation of doped ice VI samples 0.01 mol L$^{-1}$ HCl (DCl) solutions in $H_2O$ ($D_2O$) were prepared. 800 μL of these, or of pure $H_2O$ or $D_2O$ were then pipetted into indium gaskets inside the 13 mm diameter bore of a piston cylinder apparatus which was precooled to 77 K. The resulting ice I$h$ samples were then isobarically heated at 1.0 GPa to 260 K to form ice VI, quenched with liquid nitrogen while still under pressure at ~40 K min$^{-1}$, decompressed to ambient pressure, recovered under liquid nitrogen and freed from the indium. The successful preparation of ice VI was confirmed with low-temperature powder X-ray diffraction on a Stoe Stadi-P diffractometer (Cu K$_\alpha$) with a Mythen area detector.

Two to four small pieces of the high-pressure ice were then transferred into stainless-steel capsules with screwable lids under liquid nitrogen. These were quickly transferred into a precooled Perkin Elmer DSC 8000 advanced double-furnace differential scanning calorimeter. Thermograms were then recorded at different heating / cooling rates. The moles of ice in the DSC capsules were then determined by recording the endothermic melting at 0°C, and using 6012 J mol$^{-1}$ and 6280 J mol$^{-1}$ as the enthalpies of melting of H$_2$O and D$_2$O ice I$h$, respectively. The previous heating / cooling steps were then repeated with ice I$h$ and these thermograms were subtracted from the previously recorded data as a background correction. The resulting DSC signal was divided by the number of moles and the heating / cooling rate. The resulting quantity has a unit of J mol$^{-1}$ K$^{-1}$ which allows for convenient comparison between different samples and different heating / cooling rates.

## 3. Results and Discussion

Pure H$_2$O and D$_2$O ice VI show glass-transition like endothermic steps upon heating as shown in Figure 2 with onsets at ~132 and ~138 K, respectively. Such increases in heat capacity with step-heights around 1 J mol$^{-1}$ K$^{-1}$ have been observed for other hydrogen-disordered phases of ice as well, and they correspond to the kinetic unfreezing of the molecular reorientation dynamics.[19-21] The endothermic steps signal a transition from an orientational glass to a state where molecular reorientation takes place on the experimental timescale. In the case of ice VI the endothermic step is followed by a small 'plateau' region before the irreversible phase transition to stacking disordered ice (ice I*sd*) sets in.[22]

The HCl/DCl-doped and pressure-quenched samples display a very different thermal behaviour. In a quite spectacular fashion, the HCl-doped sample first displays an exotherm upon heating (–64.4 J mol$^{-1}$) followed immediately by an asymmetric endotherm (+112.0 J mol$^{-1}$) before reaching the 'plateau' region. This behaviour is explained in terms of an initial exothermic hydrogen ordering process followed by endothermic hydrogen disordering to ice VI. The pressure-quenched samples represent high free energy states at low temperatures with respect to the hydrogen-ordered phase. This is the driving force for the initial ordering process upon heating which is reverted as the hydrogen-disordered phase becomes thermodynamically more stable than the hydrogen-ordered state. The initial hydrogen ordering process is therefore by definition irreversible as it takes place from a metastable state. Using neutron diffraction it has been asserted that fully hydrogen-disordered ice VI is reached before the phase transition to ice I*sd*.[7]

In the case of the DCl-doped sample the initial hydrogen-ordering process is not observed, only the endotherm associated with hydrogen disordering. As expected, this indicates that hydrogen ordering is faster in $H_2O$ than in $D_2O$ ice VI. Furthermore, the area of the $D_2O$ endotherm (+23.1 J mol$^{-1}$) is weaker compared to the $H_2O$ endotherm. This is consistent with the fact that the $H_2O$ sample has undergone significant hydrogen ordering before reaching the temperature where hydrogen disordering sets in. This is to our knowledge the first example where such extensive 'pre-ordering' has been observed upon heating ice.

We next demonstrate the reversibility of the ice VI/XV phase transition at ambient pressure. For this, the HCl and DCl-doped samples are cooled from 138 K at 5 K min$^{-1}$ and then reheated at 5 K min$^{-1}$. The resulting thermograms are labelled as (1) and (2) in Figure 3. Exothermic hydrogen ordering is observed upon cooling and the phase transition is essentially 'mirrored' with some hysteresis upon subsequent heating.

The shape of the H$_2$O exotherm gives insights into the kinetics of the hydrogen ordering phase transition. A relatively sharp component is observed first indicating a fast initial hydrogen ordering process. This is followed by a broad 'tail' which covers a temperature range of about 30 degrees. For the D$_2$O sample, the 'tail' is not as pronounced indicating that the hydrogen ordering process is frozen-in at higher temperatures upon cooling at 5 K min$^{-1}$.

To further study the kinetics of the hydrogen ordering phase transitions the HCl and DCl-doped samples were cooled from 138 K at 30 and 0.5 K min$^{-1}$, and again reheated at 5 K min$^{-1}$. The corresponding thermograms are labelled (3) and (4) in Figure 3, and the enthalpies of all ice XV → ice VI phase transitions recorded upon heating are plotted in Figure 4(a) as a function of the cooling rate during hydrogen ordering.

A concern with these studies was that spending time at 138 K to start the cooling steps could lead to partial conversion of the samples to ice I$sd$ which would then have an impact on the recorded ice XV → ice VI transition enthalpies. To exclude this possibility the enthalpy of the ice VI → ice I$sd$ phase transitions were determined for all samples and found to be –1523 ± 11 J mol$^{-1}$ for H$_2$O ice VI. This value is even more exothermic than the previously reported value of –1404 ± 20 J mol$^{-1}$ by Handa *et al.*[23] We can therefore conclude that at least not significant amounts of the ice VI have converted to ice I$sd$ at 138 K. For D$_2$O ice VI the transition enthalpies to ice I$sd$ have not yet been reported in the literature. Yet, the onset temperature for the D$_2$O phase transition to ice I$sd$ is higher than for H$_2$O as it can be seen in Figures 1 and 2. The phase transition of D$_2$O ice VI to ice I$sd$ is therefore less likely to occur at 138 K compared to the H$_2$O phase. The D$_2$O ice VI to ice I$sd$ transition enthalpy was determined as –1561 ± 22 J mol$^{-1}$ which is, consistent with calorimetric data of other high-pressure ices,[20] more exothermic than the corresponding H$_2$O transition enthalpy.

Since we have demonstrated the reversibility of the ice VI ↔ ice XV phase transition at ambient pressure it is possible to calculate the entropy changes associated with the phase transitions from the transition enthalpies. This is achieved by dividing the DSC signal by the temperature before performing the integration. For the ice XV/VI thermograms this means that the heat released in the 'tail' region of the exotherm contributes more strongly to changes in entropy compared to the sharp heat release at higher temperatures.

Pauling estimated the molar configurational entropy of a completely hydrogen-disordered phase of ice with no ring structures as $R \ln(3/2)$.[24] Within experimental errors this value was confirmed later for ice I$h$.[25] According to the third law of thermodynamics the configurational entropy of a completely hydrogen-ordered phase of ice is zero. The Pauling entropy therefore also reflects the maximal entropy change expected for a phase transition from complete hydrogen order to complete hydrogen disorder. In Figure 4(b) we show the experimental transition entropies as fractions of the Pauling entropy. Since the ice VI state above the hydrogen ordering temperature has been shown to be fully hydrogen-disordered[7] the deviations from unity in Figure 4(b) can be taken as a measure for the residual hydrogen disorder present in the ice XV samples. It should be noted at this point that there is no linear relationship between the residual entropy and the fractional occupancies of the hydrogen sites as determined by neutron diffraction. For a hydrogen ordering process that can be described by a single ordering parameter the fractional occupancies of the hydrogen sites change from ½ to 0.84 and 0.16 upon going from complete hydrogen disorder to a state with 50% of the Pauling entropy as the residual entropy.[26]

On the basis of the data shown in Figure 4(b) the following conclusions can be made: (i) The pressure-quenching of the HCl and DCl-doped samples leads to very similar states which are close to fully hydrogen-disordered ice VI. (ii) HCl doping is effective in enabling the hydrogen ordering phase transition in $H_2O$ ice VI at ambient pressure. Increasing the cooling rate from 0.5 to 30 K min$^{-1}$ leads to only a small increase in the residual hydrogen disorder. (iii) Hydrogen ordering in DCl-doped $D_2O$ ice IV is more difficult in comparison. The hydrogen ordering phase transition can be quenched significantly upon cooling at 30 K min$^{-1}$ at ambient pressure. The slower hydrogen ordering in $D_2O$ compared to $H_2O$ is consistent with dielectric relaxation times which are generally much longer for $D_2O$ than for $H_2O$.[27,28] (iv) The gap between the $H_2O$ and $D_2O$ data points in Figure 4(b) decreases as the cooling rate is reduced from 30 to 0.5 K min$^{-1}$. This indicates that the process enabled by HCl/DCl doping is close to reaching its limits. Even slower cooling, which would be difficult experimentally, is unlikely to lead to substantial further decreases in the residual entropy. It seems therefore unlikely that the fully hydrogen-ordered state can be reached by cooling at ambient pressure. (v) Apart from having an effective dopant available there must be an additional obstacle in achieving completely hydrogen ordered ice XV.

In the case of the ice XII → XIV phase transition it has been suggested that increasing build-up of strain during the hydrogen ordering transition at ambient pressure slows down the hydrogen ordering processes.[29,30] The origins of this phenomenon are changes in the lattice parameters during the hydrogen ordering transition. However, such a phenomenon should lead to asymmetric broadening of Bragg peaks which has not been observed for $D_2O$ ice XV.[7]

Del Ben *et al.* very recently made the interesting suggestion that the dielectric properties of the surrounding medium could have an influence on the hydrogen ordering of a given ice VI crystal or domain.[17] This effect is expected to not interfere with the formation of antiferroelectric domains. Yet, they argued that the formation of ferroelectric domains should be easier in a matrix of predominately hydrogen-disordered ice VI which has a large dielectric constant.[31,32] Progressive ordering will lead to an overall decrease of the dielectric constant of a sample which could then make ferroelectric ordering difficult. Very similar arguments have very recently been put forward by Iitaka as well.[33]

As it can be seen in Figure 3 the ice VI → ice XV hydrogen ordering phase transition does not appear to be a simple one-step process. Instead, it is intriguing to speculate that the initial sharp exotherm is due to ferroelectric ordering which may be superseded by or reverted to antiferroelectric ordering at lower temperatures. In a sense, this is equivalent to saying that the intra-network hydrogen ordering takes place first followed by inter-network ordering at lower temperatures. Along these lines, the complex hydrogen ordering behaviour of ice VI could simply stem from the fact that ice VI consists of two independent networks which are not hydrogen bonded to one another. Migrating point defects, which are thought to be responsible for the hydrogen ordering, are very unlikely to 'jump' from one network to the other. The two networks should therefore have some freedom, at least during the early stages of the phase transition, to hydrogen order independently. Consistent with this scenario dielectric measurements by Johari *et al.* on pure ice VI suggested that the onset of hydrogen ordering is ferroelectric.[32,34]

It is clear that further experimental and computational work is needed in order to fully understand the complexity of the ice VI to ice XV phase transition. Previous computational studies have focussed solely on the fully hydrogen-ordered end states. However, new insights could probably be gained by studying the details of the mechanism of the phase transition considering partially hydrogen-ordered intermediate states.

**Acknowledgements**

We thank the Royal Society for a University Research Fellowship (CGS, UF100144), the Leverhulme Trust for a Research Grant (RPG-2014-04), and Dr Ben Slater for helpful discussions.

**Figures (all single column):**

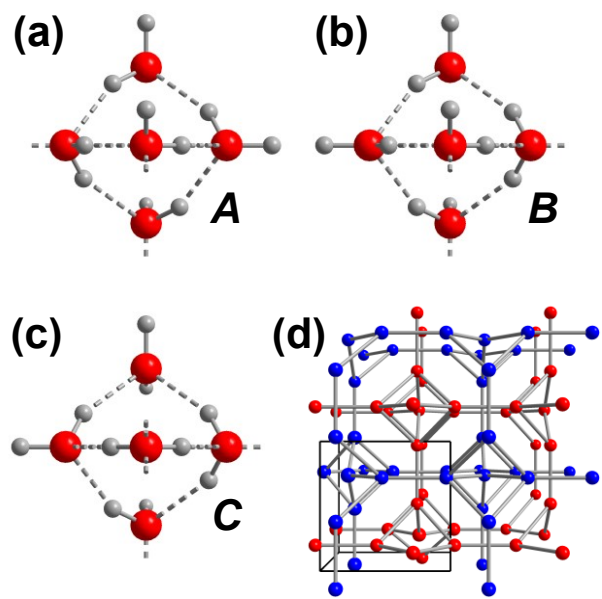

**Figure 1**. (a-c) The three different possible hydrogen-ordered structures of a single network in ice XV. (d) The packing of the networks in ice XV/VI. The oxygen atoms of the two networks are represented by red and blue spheres.

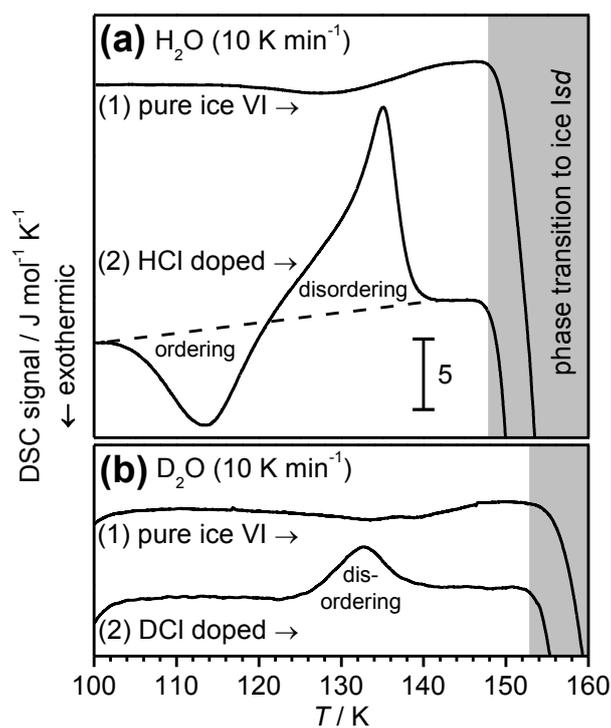

**Figure 2**. DSC thermograms recorded upon heating at 10 K min$^{-1}$ of pressure-quenched (a) pure H$_2$O and HCl-doped, and (b) pure D$_2$O and DCl-doped ice VI/XV samples. The grey-shaded regions indicate the temperature range of the irreversible phase transition to ice I$sd$. The scale bar in (a) is valid for both panels.

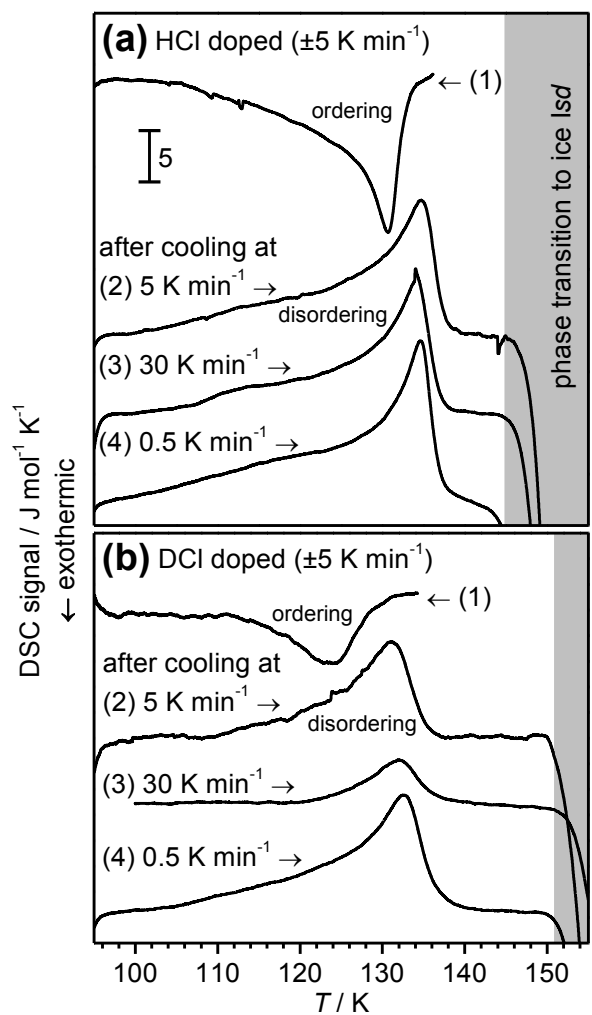

**Figure 3.** DSC thermograms of (a) HCl-doped and (b) DCl-doped ice VI/XV samples. Scans labelled with (1) were recorded upon cooling at 5 K min$^{-1}$ whereas scans (2-4) were recorded upon heating at 5 K min$^{-1}$ after previous cooling steps at 5, 30 and 0.5 K min$^{-1}$, respectively. The arrows in (a) and (b) indicate the direction of heating or cooling. The scale bar in (a) is valid for both panels.

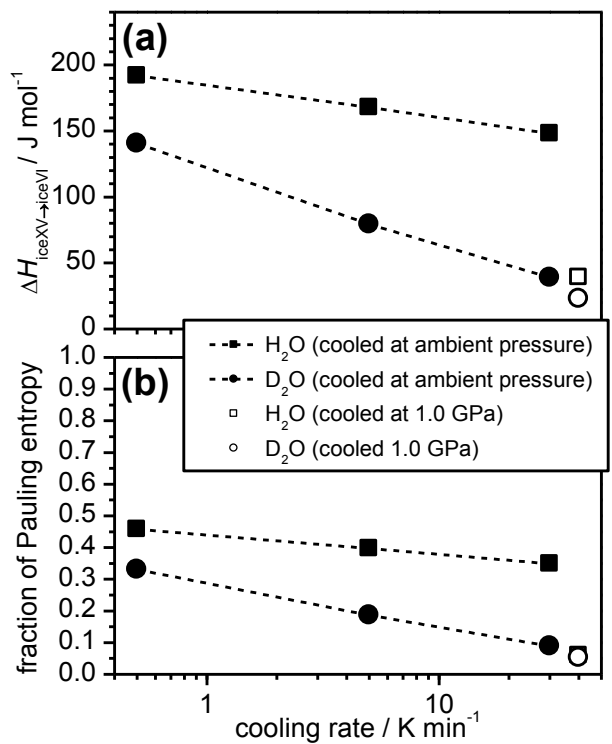

**Figure 4.** (a) Enthalpies of the ice XV → ice VI phase transitions recorded upon heating. (b) Fractions of the Pauling entropy (3.371 J mol$^{-1}$ K$^{-1}$) of the thermal features.